\documentclass[12pt]{article}
\usepackage{amssymb,amsmath,cite}

\def\nn{\nonumber}

\begin{document}

\begin{center}
{\large\bf Reduced Wigner coefficients for 
Lie superalgebra $gl(m|n)$
corresponding to unitary representations and beyond}\\
~~\\

{\large Jason L. Werry, Phillip S. Isaac and Mark D. Gould}\\
~~\\

School of Mathematics and Physics, The University of Queensland, St Lucia QLD 4072, Australia.
\end{center}

\begin{abstract}
In this paper fundamental Wigner coefficients are determined
algebraically by considering the eigenvalues of certain generalized Casimir invariants. Here this
method is applied in the context of both type 1 and type 2 unitary representations of the Lie superalgebra $gl(m|n)$. Extensions to the non-unitary case are investigated. A symmetry relation between two classes of Wigner coefficients is given in terms of a ratio of dimensions.
\end{abstract}

%

\section{Introduction}
The application of the theory of Lie Superalgebras to problems in mathematical physics relies on the
construction of an explicit set of basis states of an irreducible representation and also the
explicit determination of Wigner (Clebsch-Gordan) coefficients. Prior to the development of Lie
superalgebra theory, the foundational papers by Gel'fand and Tsetlin \cite{GT1950,GT1950b} gave a
novel construction of basis vectors for the irreducible representations of the unitary and
orthogonal groups. Further work by Baird and Biedenharn \cite{BB1963} gave a proof of Gel'fand and
Tsetlin's results while also giving, for the first time, both the fundamental and reduced Wigner
coefficients for the Lie group $U(n)$. In the same paper, the factorization of a matrix element into a
Wigner coefficient and a reduced matrix element was also examined. For a discussion on the
importance of Wigner coefficients in applications in physics, see \cite{BieLou1981}.

Lie superalgebras arose from the development of generalised Fermi-Bose statistics 
within elementary particle physics.  Physical applications of this theory have found use 
within such areas as supersymmetric integrable models 
\cite{GalMar2004,EssFraSal2005,ZhYaZh2006,FraMar2011},
logarithmic conformal field theories \cite{SchomSal2006,Ridout2009} 
and nuclear physics \cite{Iachello1988}. 

The utility of the characteristic identities (polynomial identities satisfied by generators of the
Lie algebra) became apparent during the 1970s and 80s and resulted in the
algebraic determination of reduced matrix elements \cite{Gould1978}, raising and lowering generators
\cite{Gould1980} and matrix elements \cite{Gould1981,Gould1981b}. In the series of papers
\cite{GIW1,GIW2,GIW3}
formulae for the matrix elements of unitary representations of $gl(m|n)$ were given explicitly using similar
techniques. The resulting closed-form expressions were obtained by utilizing the factorization of a
matrix element into a Wigner coefficient and a reduced matrix element. The vector Wigner
coefficients thus obtained allow us in this paper to obtain all fundamental $gl(m|n)$ Wigner
coefficients (WCs) in the Gelfand-Tsetlin (GT) basis for both type 1 and type 2 unitary
representations. Although results for vector coefficients of type 1 unitary representations have 
previously appeared in \cite{StoiVan2010}, in this presentation we also obtain results for type 2 unitary representations
and beyond. In addition, we utilize algebraic methods that can be directly generalised to other Lie
superalgebras as well as the quantum case. Furthermore, in section \ref{Extentions} we show that the
two possible forms under consideration are essentially equivalent.
Together with a continuity argument and a remarkable symmetry relation, these investigations allow 
the generalization of our results to the non-unitary case and we explicitly give Wigner coefficients 
in a non-unitary setting for the first time. 


\section{Characteristic identities and associated invariants}
\label{prelim}

We utilise the notation used in the series \cite{GIW1,GIW2,GIW3}. The generators of the Lie superalgebra $gl(m|n)$ are denoted $E_{pq}$ where $1 \leq p,q \leq m+n$. The values $1 \leq p,q \leq m$ are called \textit{even} while the values $m < p,q \leq m+n$ are called \textit{odd}.

The graded index notation will be used where the Latin indices $i,j,...$ are used for even quantities and the Greek indices $\mu, \nu, ...$ for odd quantities. The grading operator $(~)$ will then give
$$
(i) = 0,~(\mu) = 1
$$
and the full set of generators is then given by 
$$
E_{ij}, E_{i\mu}, E_{\mu i}, E_{\mu\nu}.
$$
We reserve the indices $p,q,r$ to be ungraded and to range fully from $1$ to $m+n$.

A weight $\Lambda$ may be expanded in terms
of the elementary weights $\varepsilon_p$ ($1\leq p\leq m+n$) so that
\begin{equation*}
\Lambda = \sum_{p=1}^{m+n}\Lambda_p\varepsilon_p 
\end{equation*}
where $\varepsilon_p$ is the $m+n$-tuple with $1$ in position $p$ and zeros elsewhere.

The adjoint matrix $\bar{\mathcal{A}}$ constructed in \cite{GIW1} plays an important role
in what follows and is defined as the $m+n$ square matrix with entries
$$
\bar{\mathcal{A}}_{pq} = -(-1)^{(p)(q)} E_{qp}.
$$
When $\bar{\mathcal{A}}$ acts on an irreducible $gl(m|n)$ module $V(\Lambda)$ of highest weight $\Lambda$ it will satisfy the characteristic identity
\begin{align}
\prod_{p=1}^{m+n} ({\cal \bar{A}} - \bar{\alpha}_p) = 0
\label{CharIdent1}
\end{align}
where the adjoint roots $\bar{\alpha}_i, \bar{\alpha}_\mu$ are given in terms of the highest weight labels
$$
\Lambda = (\Lambda_{i=1},...,\Lambda_{i=m}|\Lambda_{\mu=1},...\Lambda_{\mu=n})
$$
as
\begin{align*}
{\bar{\alpha}}_i &= i - 1 -\Lambda_i  , &1\leq i\leq m, \\
{\bar{\alpha}}_\mu &= \Lambda_\mu + m + 1 - \mu,  &1\leq \mu \leq n. 
\end{align*}
Immediately from the characteristic identity, we see that for each integer $r$ where $1 \leq  r \leq m+n$ there exist the projection operators 
\begin{align}
\label{proj_bar}
\bar{P}[r] = \prod^{m+n}_{p \neq r} \left( \frac{{\cal \bar{A}} - \bar{\alpha}_p}{\bar{\alpha}_r
- \bar{\alpha}_p} \right).
\end{align}
Similarly we have the vector matrix $\mathcal{A}$ with entries
$$
\mathcal{A}_{pq} = -(-1)^{(p)} E_{pq}
$$
that satisfy the polynomial identities on an irreducible $gl(m|n)$ module
\begin{align}
\prod_{p=1}^{m+n} ({\cal A} - \alpha_p) = 0
\label{CharIdent2}
\end{align}
where
\begin{align*}
{\alpha}_i &= \Lambda_i + m - n - i , &1\leq i\leq m, \\
{\alpha}_\mu &= \mu - \Lambda_\mu -n ,  &1\leq \mu \leq n. 
\end{align*}
The associated projection operators 
are then given by
\begin{align}
P[r] = \prod^{m+n}_{p \neq r} \left( \frac{{\cal A} - \alpha_p}{\alpha_r
- \alpha_p} \right).
\label{proj_unbar}
\end{align}

We now define the $gl(m|n-1)$ analogue of the matrix $\bar{\mathcal{A}}_{pq}$ so that we have the corresponding $(m+n-1) \times (m+n-1)$ square matrix
$$
\bar{\mathcal{A}_0}_{pq} = -(-1)^{(p)(q)} E_{qp}
$$
that satisfies the usual polynomial identities on an irreducible $gl(m|n-1)$ module with highest weight $\lambda$:
\begin{align*}
\prod_{p=1}^{m+n-1} (\bar{\mathcal{A}_0} - \bar{\alpha_0}_p) = 0.
\end{align*}
Here the roots are given by
\begin{align*}
\bar{\alpha_0}_i  &= i - 1 -{\lambda}_i , &1\leq i\leq m, \\
\bar{\alpha_0}_\mu &= {\lambda}_\mu + m + 1 - \mu,  &1\leq \mu \leq n - 1,
\end{align*}
and the $gl(m|n-1)$ projection operators are given by
\begin{align}
\bar{P_0}[r] = \prod_{p\neq r}^{m+n-1}\left( \frac{\bar{\mathcal{A}_0}  -\bar{\alpha_0}_p}{\bar{\alpha_0}_r - \bar{\alpha_0}_p} \right).
\end{align}
The betweenness conditions imply 
\cite{GIW1}, for $1\leq i\leq m$, that we have only two cases
\begin{align}
\bar{\alpha}_i = \left\{ \begin{array}{rl}  \bar{\alpha_0}_i - 1,& \Lambda_i = 1+\lambda_i\\
                                    \bar{\alpha_0}_i,& \Lambda_i = \lambda_i  ~~.
\end{array} \right.
\nn
\end{align}
We therefore define the following index sets
\begin{align}
I_0 &=  \{ 1\leq i\leq m\ |\ \bar{\alpha_0}_i=1+\bar{\alpha}_i \},\nn\\
\bar{I}_0 &=  \{ 1\leq i\leq m\ |\ \bar{\alpha_0}_i=\bar{\alpha}_i  \},\nn\\
I_1 &= \{ 1\leq\mu\leq n-1\},\nn\\
I &= I_0\cup I_1,\nn\\
I'&= \bar{I}_0\cup I_1,\nn\\
\tilde{I} &= I\cup \{m+n\},\nn\\
\tilde{I}' &= I'\cup \{m+n\}.
\label{DefIndexSets1} 
\end{align}
The operators
\begin{align*}
\bar{c}_r = \bar{P}[r]_{m+n}^{\ m+n}
\end{align*}
are essentially squares of reduced Wigner coefficients and this correspondence will be explained more fully in the next section. By considering the characteristic identity (\ref{CharIdent1}) satisfied by the matrix $\bar{\mathcal{A}}$ we may obtain the invariant $\bar{c}_r$ as a rational polynomial in terms of the roots $\bar{\alpha}_i$ and $\bar{\alpha}_\mu$. Specifically, we have \cite{GIW1}
\begin{align}
\bar{c}_r = \prod_{k\in \tilde{I}',k\neq r} \left(\bar{\alpha}_r - \bar{\alpha}_k\right)^{-1}\prod_{k\in
I'} \left(\bar{\alpha}_r - \bar{\alpha_0}_k - (-1)^{(k)}\right),\ \ r\in \tilde{I}',
\label{barseer}
\end{align}

Note that equation (\ref{barseer}) is positive as expected. The definition of the projection $\bar{P_0}[r]$ allows the calculation of the invariant $\bar{\rho}_{ru}$ in the following expressions \cite{GIW2}
\begin{align}
(\bar{P}_0[u]\bar{P}[r]\bar{P}_0[u])_p^{\ q} &= \bar{\rho}_{ru}\bar{P}_0[u]_p^{\ q}.
\end{align}
Explicitly, $\bar{\rho}_{ru}$ is the $gl(m|n-1)$ invariant operator given by
\begin{align}
\bar{\rho}_{ru} &= (\bar{\alpha}_r-\bar{\alpha_0}_u +
1)^{-1}(\bar{\alpha}_r-\bar{\alpha_0}_u)^{-1}\bar{c}_r\bar{\delta}_u, \ \ \ (u) = 1, \nn\\
\bar{\rho}_{ru} &= (\bar{\alpha}_r-\bar{\alpha_0}_u +
1)^{-1}(\bar{\alpha}_r-\bar{\alpha_0}_u)^{-1}\bar{c}_r\bar{\delta}_u \ \ \ (u) = 0, u \neq r,
\nn\\
\bar{\rho}_{uu} &= \bar{c}_u \bar{\delta}_u, \ \ \ (u) = 0, \nn\\
\bar{\rho}_{ru} &= (\bar{\alpha}_r-\bar{\alpha_0}_u -
1)^{-1}(\bar{\alpha}_r-\bar{\alpha_0}_u)^{-1}\bar{c}_r \bar{\delta}_u \ \ \ gl(m)
\hbox{~case,} 
\label{RhoBarExp}
\end{align}
where
\begin{align}
\bar{\delta}_u &= (-1)^{|I'|} \prod_{k\in I',k\neq u}\left(\bar{\alpha_0}_u - \bar{\alpha_0}_k -
(-1)^{(k)}\right)^{-1} \prod_{k\in \tilde{I}'}\left(\bar{\alpha}_k - \bar{\alpha_0}_u \right), u\in I, 
\end{align}
are the reduced matrix elements.
Note that $\bar{\rho}_{ru}$ in equation (\ref{RhoBarExp}) is non-vanishing only when $r\in \tilde{I}'$ and $u\in I$.

Substituting the expressions for $\bar{c}_r$ and $\bar{\delta}_u$ into equation 
 (\ref{RhoBarExp}) gives
\begin{align*}
\bar{\rho}_{ru} &= (\bar{\alpha}_r-\bar{\alpha_0}_u + 1)^{-1}\prod_{k\in \tilde{I'},k\neq r} \left(\bar{\alpha}_r - \bar{\alpha}_k \right)^{-1}\prod_{k\in
I'} \left(\bar{\alpha}_r - \bar{\alpha_0}_k - (-1)^{(k)}\right) \\
& ~\times (-1)^{|I'|} \prod_{k\in I',k\neq u} \left(\bar{\alpha_0}_u - \bar{\alpha_0}_k -
(-1)^{(k)}\right)^{-1}\prod_{k\in\tilde{I'},k \neq r} \left(\bar{\alpha}_k - \bar{\alpha_0}_u \right),\ \ r\in \tilde{I'},u\in I_0 
\end{align*}
for $u$ even and
\begin{align*}
\bar{\rho}_{ru} &= (-1)^{|I'|} \prod_{k\in \tilde{I'},k\neq r} 
\left(
\frac 
{\bar{\alpha}_k - \bar{\alpha_0}_u  }
{\bar{\alpha}_r - \bar{\alpha}_k }
\right)
\prod_{k\in I',k \neq u} 
\left(
\frac
{ \bar{\alpha}_r - \bar{\alpha_0}_k - (-1)^{(k)} }
{\bar{\alpha_0}_u - \bar{\alpha_0}_k - (-1)^{(k)} }
\right)
,\ \ r\in \tilde{I'},u\in I_1
\end{align*}
for $u$ odd. Note that the expressions for $\bar{\rho}_{ru}$ always evaluate to positive values as expected.


Similarly, the $gl(m|n-1)$ analogue of ${\mathcal{A}}_{pq}$ is the $(m+n-1) \times (m+n-1)$ square matrix
$$
{\mathcal{A}_0}_{pq} = -(-1)^{(p)} E_{pq}
$$
that satisfies the usual polynomial identities on an irreducible $gl(m|n-1)$ module with highest weight $\lambda$ is
\begin{align*}
	\prod^{m+n-1}_{p=1} (\mathcal{A}_0 - {\alpha_0}_p ) = 0 .
\end{align*}
Here the roots are given by
\begin{align*}
{\alpha_0}_i &= {\lambda}_i + m - n + 1 - i, &1\leq i\leq m, \\
{\alpha_0}_\mu &= \mu-{\lambda}_\mu - n + 1, &1\leq \mu \leq n - 1 
\end{align*}
and the dual $gl(m|n-1)$ projection operators are expressed as
\begin{align}
P_0[r] = \prod_{p\neq r}^{m+n-1}\left( 
\frac{\mathcal{A}_0-{\alpha_0}_p}
{{\alpha_0}_r-{\alpha_0}_p} \right).
\end{align}
The betweenness conditions imply 
\cite{GIW3}, for $1\leq i\leq m$, that we have only two cases
\begin{align}
\alpha_i = \left\{ \begin{array}{rl} {\alpha_0}_i,& \Lambda_i = 1+\lambda_i\\
                                    {\alpha_0}_i - 1,& \Lambda_i = \lambda_i  ~~.
\end{array} \right.
\nn 
\end{align}
The index sets in terms of the adjoint roots $\alpha_i$ and ${\alpha_0}_i$ are then
\begin{align}
I_0 &=  \{ 1\leq i\leq m\ |\ {\alpha_0}_i=\alpha_i\},\nn\\
\bar{I}_0 &=  \{ 1\leq i\leq m\ |\ {\alpha_0}_i=1+\alpha_i\},\nn\\
I_1 &= \{ 1\leq\mu\leq n-1\},\nn\\
I &= I_0\cup I_1,\nn\\
I'&= \bar{I}_0\cup I_1,\nn\\
\tilde{I} &= I\cup \{m+n\},\nn\\
\tilde{I}' &= I'\cup \{m+n\}.
\label{DefIndexSets2} 
\end{align}
Similarly, we define the analogue of the $\bar{c}_r$ operator 
\begin{align*}
c_r = P[r]_{\ m+n}^{m+n}
\end{align*}
which is also related to the squares of reduced Wigner coefficients.

By considering the characteristic identities satisfied by the matrix $\bar{\mathcal{A}}$ we may obtain the invariant $c_r$ as a rational polynomial in terms of the roots $\alpha_i,\alpha_\mu$ and ${\alpha_0}_i,{\alpha_0}_\mu$. Specifically, we have \cite{GIW1}
\begin{align}
c_r = \prod_{k\in \tilde{I},k\neq r} \left(\alpha_r - \alpha_k \right)^{-1}\prod_{k\in
I} \left(\alpha_r - {\alpha_0}_k - (-1)^{(k)}\right),\ \ r\in \tilde{I} ,
\label{seer}
\end{align}
and note that equation (\ref{seer}) is positive (as expected).

The definitions of the projections ${P_0}[r]$ and ${P}[r]$ allow the calculation of the invariant ${\rho}_{ru}$ in the following expressions \cite{GIW2}
\begin{align}
({P_0}[u]{P}[r]{P_0}[u])_p^{\ q} &= {\rho}_{ru}{P_0}[u]_{\ p}^q 
\end{align}
where ${\rho}_{ru}$ are the $gl(m|n-1)$ invariant operators given by
\begin{align}
{\rho}_{ru} &= ({\alpha}_r-{\alpha_0}_u +
1)^{-1}({\alpha}_r-{\alpha_0}_u)^{-1}{c}_r{\delta}_u, \ \ \ (u) = 1, \nn\\
{\rho}_{ru} &= ({\alpha}_r-{\alpha_0}_u +
1)^{-1}({\alpha}_r-{\alpha_0}_u)^{-1}{c}_r{\delta}_u \ \ \ (u) = 0, u \neq r,
\nn\\
{\rho}_{uu} &= {c}_u {\delta}_u, \ \ \ (u) = 0, \nn\\
{\rho}_{ru} &= ({\alpha}_r-{\alpha_0}_u -
1)^{-1}({\alpha}_r-{\alpha_0}_u)^{-1}{c}_r {\delta}_u \ \ \ gl(m)
\hbox{~case,} 
\label{RhoExp}
\end{align}
where 
\begin{align}
\delta_u &= (-1)^{|I|} \prod_{k\in I,k\neq u} \left({\alpha_0}_u - {\alpha_0}_k -
(-1)^{(k)}\right)^{-1}\prod_{k\in\tilde{I}} \left(\alpha_k - {\alpha_0}_u \right),\ \ u\in I',
\end{align}
are the reduced matrix elements.
Note that ${\rho}_{ru}$ in equation (\ref{RhoExp}) is non-vanishing only when $r\in \tilde{I}$ and $u\in I'$.

Substituting the expressions for ${c}_r$ and ${\delta}_u$ into equation 
 (\ref{RhoExp}) initially gives
\begin{align*}
{\rho}_{ru} &= ({\alpha}_r-{\alpha_0}_u + 1)^{-1}({\alpha}_r-{\alpha_0}_u)^{-1} \prod_{k\in \tilde{I},k\neq r} \left(\alpha_r - \alpha_k \right)^{-1}\prod_{k\in
I} \left(\alpha_r - {\alpha_0}_k - (-1)^{(k)}\right) \\
& ~\times (-1)^{|I|} \prod_{k\in I,k\neq u} \left({\alpha_0}_u - {\alpha_0}_k -
(-1)^{(k)}\right)^{-1}\prod_{k\in\tilde{I}} \left(\alpha_k - {\alpha_0}_u \right),\ \ r\in \tilde{I},u\in I' .
\end{align*}
Now $({\alpha}_r-{\alpha_0}_u)^{-1}$ will cancel the corresponding term from $\delta_u$ while $
({\alpha}_r-{\alpha_0}_u + 1)^{-1}$ will only cancel a term in $c_r$ when $u$ is odd. We therefore have
\begin{align*}
{\rho}_{ru} &= ({\alpha}_r-{\alpha_0}_u + 1)^{-1}\prod_{k\in \tilde{I},k\neq r} \left(\alpha_r - \alpha_k \right)^{-1}\prod_{k\in
I} \left(\alpha_r - {\alpha_0}_k - (-1)^{(k)}\right) \\
& ~\times (-1)^{|I|} \prod_{k\in I,k\neq u} \left({\alpha_0}_u - {\alpha_0}_k -
(-1)^{(k)}\right)^{-1}\prod_{k\in\tilde{I},k \neq r} \left(\alpha_k - {\alpha_0}_u \right),\ \ r\in \tilde{I},u\in \bar{I}_0 
\end{align*}
for $u$ even and
\begin{align*}
{\rho}_{ru} &= (-1)^{|I|} \prod_{k\in \tilde{I},k\neq r} 
\left(
\frac 
{\alpha_k - {\alpha_0}_u  }
{\alpha_r - \alpha_k }
\right)
\prod_{k\in I,k \neq u} 
\left(
\frac
{ \alpha_r - {\alpha_0}_k - (-1)^{(k)} }
{{\alpha_0}_u - {\alpha_0}_k - (-1)^{(k)} }
\right)
,\ \ r\in \tilde{I},u\in I_1
\end{align*}
for $u$ odd.
Note that the expression for $\rho_{ru}$ always evaluates to a positive value as expected.

\section{Reduced Wigner coefficients} 
\label{Wigner}

We will now show that the eigenvalues of the invariants $\bar{c}_r$ and $c_r$ are 
essentially squares of reduced Wigner coefficients. Before proceeding we must make the following remarks:

{\bf Remark 1:} 
Strictly speaking we are implicitly assuming that within the $gl(m|n)$ projection operator
expressions (\ref{proj_bar}) and (\ref{proj_unbar}) we have $\alpha_r \neq \alpha_k$ and
$\bar{\alpha}_r \neq \bar{\alpha}_k$ for all $r \neq k$. Multiplicities of the roots $\alpha_r$ and
$\bar{\alpha}_r$ are related to the occurrence of atypical irreducible representations in the tensor
product of $V\otimes V(\Lambda)$ or $V^*\otimes V(\Lambda)$, where $V=V(\varepsilon_1)$ is the vector module and
$V^*$ its dual. The set of $\Lambda$ for which this happens, however, is closed in the Zariski topology
\cite{Hump1972} on $H^*$. It follows since all formulae in our results determine rational
polynomials that such formulae extend to all dominant $\Lambda$ by continuity.

{\bf Remark 2:} 
When applying the projection operators we must take care to distinguish between type 1 unitary and type 2 unitary modules. On an irreducible finite dimensional $gl(m|n)$ module $V(\Lambda)$ there exists a non-degenerate sesquilinear form 
$\langle ~|~ \rangle_\theta$ satisfying 
\begin{align}
\label{FormType1and2}
\langle E_{pq} v | w \rangle = (-1)^{(\theta - 1)[(p) + (q)]} \langle v | E_{qp} w \rangle , ~\theta  \in \{1,2\},
\end{align} 
that is unique up to scalar multiples. The module $V(\Lambda)$ is called type $\theta$ unitary if $\langle ~|~ \rangle_\theta$ is positive definite on $V(\Lambda)$ and the corresponding representation is also called type $\theta$ unitary. The induced form on the tensor product space $V(\Lambda) \otimes V(\mu)$ induces a non-degenerate inner product on the tensor product module in terms of which Wigner coefficients are defined in the usual way.

Due to these considerations we now consider the type 1 unitary and type 2 unitary cases of Wigner coefficients separately.

\subsection{Covariant vector Wigner coefficients}

Let $e_i$ denote the Gelfand-Tsetlin (GT) basis states for the (type 1 unitary) vector module $V$ and ${e^\Lambda_\beta}$ denote the (GT) basis states for the irreducible type 1 unitary module $V(\Lambda)$ of highest weight $\Lambda$. Then the matrix elements of $\bar{P}[r]$ can be given in the form
\begin{align}
\left\langle e^\Lambda_\beta | \bar{P}[r]^j_i | e^{\Lambda}_\alpha \right\rangle =
\sum_\gamma \left\langle e^\Lambda_\beta \otimes e_i | e^{\Lambda + \varepsilon_r}_\gamma \right\rangle
\left\langle e^{\Lambda+\varepsilon_r}_\gamma | e_j \otimes e^\Lambda_\alpha \right\rangle, \label{BarPij}
\end{align}
where
$$
\left\langle e^{\Lambda+\varepsilon_r}_\gamma | e_j \otimes e^\Lambda_\alpha \right\rangle,
$$
are the vector (fundamental) Wigner coefficients.

From Schur's lemma we observe that the fundamental WCs factorize as follows
\begin{align}
\left\langle\left. 
\begin{array}{c} \Lambda+\varepsilon_k\\ \lambda+\varepsilon_{0_r} \\ {[\Lambda'_0]} \end{array}
\right|\right.
\left.
e_i\otimes \begin{array}{c} \Lambda \\ \lambda \\
{[\Lambda_0]} \end{array}
\right\rangle &= 
\left\langle \left.
\begin{array}{c} \Lambda+\varepsilon_k\\ \lambda+\varepsilon_{0_r} 
 \end{array}
\right\rVert
\begin{array}{c} \varepsilon_1 \\ \varepsilon_{0_1} 
 \end{array}
; \begin{array}{c} \Lambda \\ \lambda
 \end{array}
\right\rangle 
\left\langle\left. 
\begin{array}{c} \lambda+\varepsilon_{0_r} \\ {[\Lambda'_0]} \end{array}
\right|\right.
\left.
e_i\otimes \begin{array}{c} \lambda \\
{[\Lambda_0]} \end{array}
\right\rangle,\label{Schur1}\\ 
1 < i < m+n, \nn
\end{align}
where $1 < k \leq m+n$,~$1 < r \leq m+n-1$, 
$\Lambda$ denotes the type 1 unitary highest weight of the $gl(m|n)$ module $V(\Lambda)$, $\lambda$ denotes the highest
weight of the $gl(m|n-1)$ module $V(\lambda)$ which occurs in the decomposition of $V(\Lambda)$ into
irreducible $gl(m|n-1)$ modules, and $[\Lambda_0]$
denotes the GT pattern of the $gl(m|n-2)$ subalgebra. In
addition, the first term on the right hand side of equation (\ref{Schur1}) is a reduced Wigner
coefficient (RWC) which is independent of the highest weight labels of $[\Lambda_0]$ and
$[\Lambda'_0]$.

Setting $i=m+n$ gives
\begin{align}
\left\langle\left. 
\begin{array}{c} \Lambda+\varepsilon_k\\ \lambda \\ {[\Lambda'_0]} \end{array}
\right|\right.
\left.
e_{m+n} \otimes \begin{array}{c} \Lambda \\ \lambda \\
{[\Lambda_0]} \end{array}
\right\rangle = \delta_{[\Lambda'_0][\Lambda_0]}
 \left\langle 
\begin{array}{c} \Lambda+\varepsilon_k\\ \lambda \end{array}
\right\rVert
\left.
\begin{array}{c} 
\varepsilon_1 \\ \dot{0}
\end{array}
;
\begin{array}{c} \Lambda \\ 
\lambda \end{array}
\right\rangle  , \label{WCBraKet1}
\end{align}
where $\delta$ here denotes the Kronecker delta and the RWC on the right hand side reduces to the WC in this case which is independent of $[\Lambda_0]$ and $[\Lambda'_0]$ as expected. 
The WC in equation (\ref{WCBraKet1}) is given by the eigenvalues of the invariant
$$ 
\bar{c}_r = \bar{P}[r]_{m+n}^{\ m+n}
$$
since from equation (\ref{BarPij}) we have
\begin{align}
\bar{c}_r  &= \left| \left\langle 
\begin{array}{c} \Lambda+\varepsilon_r\\ \lambda \end{array}
\right\rVert
\left.
\begin{array}{c} \varepsilon_1 \\
\dot{0} \end{array} 
;
 \begin{array}{c} \Lambda \\
\lambda \end{array}
\right\rangle \right|^2. \label{WigCoefCBar}
\end{align}

We therefore see that the matrix elements of the projection $\bar{P}[r]_{m+n}^{\ m+n}$ determine
the squares of the fundamental Wigner coefficients and depend only on the top two rows of the corresponding Gelfand-Tsetlin basis states.

Furthermore, the eigenvalues of the invariant $\bar{\rho}_{ru}$ determine the square of $gl(m|n):gl(m|n-1)$ covariant vector RWCs \cite{GIW2} since 
\begin{align}
\bar{\rho}_{ru}  &= \left| \left\langle
\begin{array}{c} \Lambda+\varepsilon_r\\ \lambda+\varepsilon_{0_u} \end{array}
\right\rVert
\left.
\begin{array}{c} \varepsilon_1 \\
\varepsilon_{0_1} \end{array} 
;
 \begin{array}{c} \Lambda \\
\lambda \end{array}
\right\rangle \right|^2.
\end{align}

The expressions for the WCs of equation (\ref{WCBraKet1}) together with the RWCs of equation (\ref{Schur1}) including their phases \cite{GIW2} are then given by
\begin{align}
& \left\langle 
\begin{array}{c} \Lambda+\varepsilon_r\\ \lambda
 \end{array}
\right\rVert
\left.
\begin{array}{c} \varepsilon_1 \\ 0 
 \end{array}
; \begin{array}{c} \Lambda \\ \lambda
 \end{array}
\right\rangle = \left[ \prod_{k\in \tilde{I}',k\neq r} \left(\bar{\alpha}_r - \bar{\alpha}_k\right)^{-1}\prod_{k\in
I'} \left(\bar{\alpha}_r - \bar{\alpha_0}_k - (-1)^{(k)}\right) \right]^{1/2}
,\ \ r\in \tilde{I}'
\end{align}

\begin{align}
& \left\langle
\begin{array}{c} \Lambda+\varepsilon_r\\ \lambda+\varepsilon_{0_u} 
 \end{array}
\right\rVert
\left.
\begin{array}{c} \varepsilon_1 \\ \varepsilon_{0_1} 
 \end{array}
; \begin{array}{c} \Lambda \\ \lambda
 \end{array}
\right\rangle = (-1)^{(r)(u)} S(r-u) \nn\\
&~~\times 
\left[ 
\frac{
(-1)^{|I'|} \prod_{k\in\tilde{I'},k \neq r} \left(\bar{\alpha}_k - \bar{\alpha_0}_u \right)
\prod_{k\in I'} \left(\bar{\alpha}_r - \bar{\alpha_0}_k - (-1)^{(k)}\right) 
}
{
(\bar{\alpha}_r-\bar{\alpha_0}_u + 1)
\prod_{k\in \tilde{I'},k\neq r} \left(\bar{\alpha}_r - \bar{\alpha}_k \right)
\prod_{k\in I',k\neq u} \left(\bar{\alpha_0}_u - \bar{\alpha_0}_k - (-1)^{(k)}\right)
}
\right]^{1/2}
, \ \ r\in \tilde{I'},u\in I_0   \label{FinalBarredEven}
\end{align}

\begin{align}
& \left\langle
\begin{array}{c} \Lambda+\varepsilon_r\\ \lambda+\varepsilon_{0_u} 
 \end{array}
\right\rVert
\left.
\begin{array}{c} \varepsilon_1 \\ \varepsilon_{0_1} 
 \end{array}
; \begin{array}{c} \Lambda \\ \lambda
 \end{array}
\right\rangle = (-1)^{(r)(u)} S(r-u) \nn\\
&~~\times \left[ (-1)^{|I'|} \prod_{k\in \tilde{I'},k\neq r} 
\left(
\frac 
{\bar{\alpha}_k - \bar{\alpha_0}_u  }
{\bar{\alpha}_r - \bar{\alpha}_k }
\right)
\prod_{k\in I',k \neq u} 
\left(
\frac
{ \bar{\alpha}_r - \bar{\alpha_0}_k - (-1)^{(k)} }
{\bar{\alpha_0}_u - \bar{\alpha_0}_k - (-1)^{(k)} }
\right)
\right]^{1/2}
,\ \ r\in \tilde{I'},u\in I_1 \label{FinalBarredOdd}
\end{align}
where the positive square root is always taken, odd indices are considered greater than even indices, 
$$
S(x) = sgn(x),~~S(0)=1,
$$
and where we recall
\begin{align*}
{\bar{\alpha_0}}_i &= i - 1 -\Lambda_i  , &1\leq i\leq m-1, \\
{\bar{\alpha_0}}_\mu &= \Lambda_\mu + m + 1 - \mu  ,  &1\leq \mu \leq n-1 , \\ 
\bar{\alpha}_i  &= i - 1 -{\tilde{\Lambda}}_i , &1\leq i\leq m, \\
\bar{\alpha}_\mu &= {\tilde{\Lambda}}_\mu + m + 1 - \mu,  &1\leq \mu \leq n.  
\end{align*}
The fundamental Wigner coefficients are given in terms of the RWCs by applying the RWC factorization (\ref{Schur1}) down the subalgebra chain:
\begin{align}
\left\langle (\Lambda') \vert e_p \otimes (\Lambda) \right\rangle 
&= \left\langle 
\begin{array}{c} \Lambda_p+\varepsilon_{r_p,p}\\ \Lambda_{p-1} \end{array}
\right\rVert
\left.
\begin{array}{c} 
\varepsilon_1 \\ \dot{0}
\end{array}
;
\begin{array}{c} \Lambda_p \\ 
\Lambda_{p-1} \end{array}
\right\rangle \nn\\
&~\times \prod_{k=p+1}^{m+n} 
\left\langle 
\begin{array}{c} \Lambda_k+\varepsilon_{r_k,k}\\ \Lambda_{k-1} +\varepsilon_{r_{k-1},k-1} \end{array}
\right\rVert
\left.
\begin{array}{c} 
\varepsilon_1 \\ \varepsilon_1
\end{array}
;
\begin{array}{c} \Lambda_k \\ 
\Lambda_{k-1} \end{array}
\right\rangle \label{FullWC1} 
\end{align}
where $(\Lambda)$ is a $gl(m|n)$ Gelfand-Tsetlin pattern with rows $\Lambda_k, 1 \leq k \leq m+n$. The pattern $(\Lambda')$ is  obtained from $(\Lambda)$ by raising the weight label $\Lambda_{r_k,k}, 1 \leq r_k \leq k$ by one unit while preserving the remaining labels of the row $\Lambda_k$.

\subsection{Contravariant dual vector Wigner coefficients}
\label{SubSect_ContraDualWig}
We now turn to the dual fundamental Wigner coefficients and proceed similarly. 
Denoting $\bar{e}_i$ to be the Gelfand-Tsetlin (GT) basis states of the (type 2 unitary) dual vector module and ${e^\Lambda_\beta}$ to be the (GT) basis states for the irreducible type 2 unitary module $V(\Lambda)$ of highest weight $\Lambda$ we have 
\begin{align}
\left\langle e^\Lambda_\beta | P[r]^j_i | e^{\Lambda}_\alpha \right\rangle =
\sum_\gamma \left\langle e^\Lambda_\beta \otimes \bar{e}_i | e^{\Lambda - \varepsilon_r}_\gamma \right\rangle
\left\langle e^{\Lambda-\varepsilon_r}_\gamma | \bar{e}_j \otimes e^\Lambda_\alpha \right\rangle, \label{Pij}
\end{align}
where
$$
\left\langle e^{\Lambda-\varepsilon_r}_\gamma | \bar{e}_j \otimes e^\Lambda_\alpha \right\rangle,
$$
are the dual fundamental Wigner coefficients. The above quantities may be expressed in terms of the $gl(m|n)$ highest weight labels of $\Lambda$ and the highest weight labels of the $gl(m|n-1)$ subalgebra.

Let $\bar{\varepsilon}_k$ denote the weights of the $gl(m|n)$ contravariant vector irrep. These weights are given by $\bar{\varepsilon}_k = -\varepsilon_{m+n+1-k} (1 \leq k \leq m+n)$. We similarly define the $gl(m|n-1)$ fundamental weights $\bar{\varepsilon}_{0_r} =
\varepsilon_{0_{m+n-r}} = (1 \leq r \leq m+n-1) $.

Again, from Schur's lemma we observe that the dual fundamental WCs factorize as follows
\begin{align}
\left\langle\left. 
\begin{array}{c} \Lambda-\varepsilon_k\\ \lambda-\varepsilon_{0_r} \\ {[\Lambda'_0]} \end{array}
\right|\right.
\left.
\bar{e}_i\otimes \begin{array}{c} \Lambda \\ \lambda \\
{[\Lambda_0]} \end{array}
\right\rangle &= 
\left\langle 
\begin{array}{c} \Lambda-\varepsilon_k\\ \lambda-\varepsilon_{0_r} 
 \end{array}
\right\rVert
\left.
\begin{array}{c} \bar{\varepsilon}_1 \\ \bar{\varepsilon}_{0_1} 
 \end{array}
; \begin{array}{c} \Lambda \\ \lambda
 \end{array}
\right\rangle 
\left\langle\left. 
\begin{array}{c} \lambda-\varepsilon_{0_r} \\ {[\Lambda'_0]} \end{array}
\right|\right.
\left.
\bar{e}_i\otimes \begin{array}{c} \lambda \\
{[\Lambda_0]} \end{array}
\right\rangle,\label{Schur2}\\
~~\nn\\
1 < i < m+n, \nn
\end{align}
where $1 < k \leq m+n$,~$1 < r \leq m+n-1$, 
$\Lambda$ denotes the type 2 unitary highest weight of the $gl(m|n)$ module $V(\Lambda)$, $\lambda$ denotes the highest
weight of the $gl(m|n-1)$ module $V(\lambda)$ which occurs in the decomposition of $V(\Lambda)$ into
irreducible $gl(m|n-1)$ modules, and $[\Lambda_0]$ denotes the GT pattern of the $gl(m|n-2)$ subalgebra. As in
the vector Wigner coefficient case, the first term on the right hand side of equation (\ref{Schur2}) is a reduced Wigner coefficient (RWC) which is independent of the highest
weight labels of $[\Lambda_0]$ and $[\Lambda'_0]$.

Setting $i=m+n$ gives
\begin{align}
\left\langle\left. 
\begin{array}{c} \Lambda-\varepsilon_k\\ \lambda \\ {[\Lambda'_0]} \end{array}
\right|\right.
\left.
\bar{e}_{m+n} \otimes \begin{array}{c} \Lambda \\ \lambda \\
{[\Lambda_0]} \end{array}
\right\rangle = \delta_{[\Lambda'_0][\Lambda]}
 \left\langle 
\begin{array}{c} \Lambda-\varepsilon_k\\ \lambda \end{array}
\right\rVert
\left.
\begin{array}{c} 
\bar{\varepsilon}_1 \\ \dot{0}
\end{array}
;
\begin{array}{c} \Lambda \\
\lambda \end{array}
\right\rangle  , \label{WCBraKet2}
\end{align}
where the RWC on the right hand side reduces to the WC in this case which are independent of $[\Lambda_0]$ and $[\Lambda'_0]$ as expected. 
The WC in equation (\ref{WCBraKet2}) is given by the eigenvalues of the invariant
$$ 
c_r = P[r]_{\ m+n}^{m+n}
$$
since from equation (\ref{Pij}) we have
\begin{align}
c_r &= \left| \left\langle
\begin{array}{c} \Lambda-\varepsilon_r\\ \lambda \end{array}
\right\rVert
\left.
\begin{array}{c} \bar{\varepsilon}_1 \\
\dot{0} \end{array}  ; \begin{array}{c} \Lambda \\
\lambda \end{array}
\right\rangle \right|^2. \label{WigCoefCUnbar}
\end{align}
As in the vector fundamental Wigner case we see that the matrix elements of the projection  $P[r]_{\ m+n}^{m+n}$ determine
the squares of the dual fundamental Wigner coefficients and depend only on the top two rows of the corresponding Gelfand-Tsetlin basis states.

In addition, the eigenvalues of the invariant ${\rho}_{ru}$ determine the square of $gl(m|n):gl(m|n-1)$ dual vector RWCs \cite{GIW2} since
\begin{align}
\rho_{ru} &= \left| \left\langle 
\begin{array}{c} \Lambda-\varepsilon_r\\ \lambda-\varepsilon_{0_u} \end{array}
\right\rVert
\left.
\begin{array}{c} \bar{\varepsilon}_1 \\
\bar{\varepsilon}_{0_1}  \end{array}  ; \begin{array}{c} \Lambda \\
\lambda \end{array}
\right\rangle \nn \right|^2.
\end{align}

The expressions for the WCs of equation (\ref{WCBraKet2}) together with the RWCs of equation (\ref{Schur2}) including their phases \cite{GIW3} are then given by
\begin{align}
&\left\langle 
\begin{array}{c} \Lambda-\varepsilon_r\\ \lambda
 \end{array}
\right\rVert
\left.
\begin{array}{c} \bar{\varepsilon}_1 \\ 0 
 \end{array}
; \begin{array}{c} \Lambda \\ \lambda
 \end{array}
\right\rangle
= \left[ \prod_{k\in \tilde{I},k\neq r} \left(\alpha_r - \alpha_k \right)^{-1}\prod_{k\in
I} \left(\alpha_r - {\alpha_0}_k - (-1)^{(k)}\right) \right]^{1/2}
,\ \ r\in \tilde{I} .
\end{align}

\begin{align}
&\left\langle
\begin{array}{c} \Lambda-\varepsilon_r\\ \lambda-\varepsilon_{0_u} 
 \end{array}
\right\rVert
\left.
\begin{array}{c} \bar{\varepsilon}_1 \\ \bar{\varepsilon}_{0_1} 
 \end{array}
; \begin{array}{c} \Lambda \\ \lambda
 \end{array}
\right\rangle
= (-1)^{(r)(u) +(r) + (u)} S(r-u)
\nn \\
&~~\times
\left[
\frac{
(-1)^{|I|}
\prod_{k\in I} \left(\alpha_r - {\alpha_0}_k - (-1)^{(k)}\right)
\prod_{k\in\tilde{I},k \neq r} \left(\alpha_k - {\alpha_0}_u \right)
}
{
({\alpha}_r-{\alpha}_u + 1)
\prod_{k\in \tilde{I},k\neq r} \left(\alpha_r - \alpha_k \right)
\prod_{k\in I,k\neq u} \left({\alpha_0}_u - {\alpha_0}_k - (-1)^{(k)}\right)
} 
\right]^{1/2}
,\ \ r\in \tilde{I},u\in \bar{I}_0 \label{FinalUnBarredEven}
\end{align}

\begin{align}
&\left\langle 
\begin{array}{c} \Lambda-\varepsilon_r\\ \lambda-\varepsilon_{0_u} 
 \end{array}
\right\rVert
\left.
\begin{array}{c} \bar{\varepsilon}_1 \\ \bar{\varepsilon}_{0_1} 
 \end{array}
; \begin{array}{c} \Lambda \\ \lambda
 \end{array}
\right\rangle = (-1)^{(r)(u) +(r) + (u)} S(r-u) 
\nn\\
&~~\times
\left[
 (-1)^{|I|} \prod_{k\in \tilde{I},k\neq r} 
\left(
\frac 
{\alpha_k - {\alpha_0}_u  }
{\alpha_r - \alpha_k }
\right)
\prod_{k\in I,k \neq u} 
\left(
\frac
{ \alpha_r - {\alpha_0}_k - (-1)^{(k)} }
{{\alpha_0}_u - {\alpha_0}_k - (-1)^{(k)} }
\right)
\right]^{1/2}
,\ \ r\in \tilde{I},u\in I_1  \label{FinalUnBarredOdd}
\end{align}
where the positive square root is always taken, odd indices are considered greater than even indices, 
$$
S(x) = sgn(x),~~S(0)=1.
$$
and where we recall
\begin{align*}
{\alpha_0}_i &= \Lambda_i + m - n + 1 - i  , &1\leq i\leq m-1, \\
{\alpha_0}_\mu &= \mu - \Lambda_\mu -n + 1, &1\leq \mu \leq n-1 , \\
\alpha_i &= {\tilde{\Lambda}}_i + m - n - i, &1\leq i\leq m, \\
\alpha_\mu &= \mu-{\tilde{\Lambda}}_\mu - n, &1\leq \mu \leq n .  
\end{align*}
The fundamental Wigner coefficients are given in terms of the RWCs via application of the factorization $(\ref{Schur2})$ as
\begin{align}
\left\langle (\Lambda') \vert \bar{e}_p \otimes (\Lambda) \right\rangle 
&= \left\langle 
\begin{array}{c} \Lambda_p-\varepsilon_{r_p,p}\\ \Lambda_{p-1} \end{array}
\right\rVert
\left.
\begin{array}{c} 
\bar{\varepsilon}_1 \\ \dot{0}
\end{array}
;
\begin{array}{c} \Lambda_p \\ 
\Lambda_{p-1} \end{array}
\right\rangle \nn\\
&~\times \prod_{k=p+1}^{m+n} 
\left\langle 
\begin{array}{c} \Lambda_k-\varepsilon_{r_k,k}\\ \Lambda_{k-1} -\varepsilon_{r_{k-1},k-1} \end{array}
\right\rVert
\left.
\begin{array}{c} 
\bar{\varepsilon}_1 \\ \bar{\varepsilon}_1
\end{array}
;
\begin{array}{c} \Lambda_k \\ 
\Lambda_{k-1} \end{array}
\right\rangle \label{FullWC2} 
\end{align}
where the notation matches that of equation (\ref{FullWC1}) except that $(\Lambda')$ is  obtained from $(\Lambda)$ by lowering the weight label $\Lambda_{r_k,k}, 1 \leq r_k \leq k$ by one unit.

\section{Extension to non-unitary modules: equivalent forms and symmetry relations}
\label{Extentions}
In this section we emphasize that even on non-unitary representations we may define vector and dual
vector Wigner coefficients with respect to the naturally induced form on an irreducible module. In
fact there are two such forms as expressed in equation (\ref{FormType1and2}). In section
\ref{ExtentionsPoint1} we show that these forms are essentially equivalent (up to a phase).
For the purposes of this section $\langle ~ | ~ \rangle_1$ is equation (\ref{FormType1and2}) with
$\theta = 1$ and $\langle ~ | ~ \rangle_2$ is equation (\ref{FormType1and2}) with $\theta = 2$.

\subsection{Connections between forms}
\label{ExtentionsPoint1}
It is instructive to consider the form in equation (\ref{FormType1and2}) under the action of the grading automorphism $\gamma$ \cite{Sch1979} given by
\begin{align*}
\gamma: L \rightarrow L = L_0 \oplus L_1 \\
\gamma(x_0 + x_1) = x_0 - x_1.
\end{align*}
where $L_0$ (resp. $L_1$) is the even (resp. odd) submodule of $L$ and $x_0,x_1$ are the homogeneous vectors with $x_0 \in L_0$, $x_1 \in L_1$.
From equation (\ref{FormType1and2}) we recall the form $\langle ~ | ~ \rangle_1$ satisfies
$$
\langle E_{pq} v | w \rangle_1 = \langle v | E_{qp} w \rangle_1
$$
and we also define the new form
$$
\langle v \vert w \rangle_1' = \langle v \vert \gamma(w)\rangle_1.
$$
From 
\begin{align*}
\langle E_{pq} v |w \rangle_1' &= \langle E_{pq} v | \gamma(w) \rangle_1 \\
&= \langle v | E_{pq} \gamma(w) \rangle_1 \\
&= (-1)^{[p] + [q]} \langle v | \gamma(E_{pq} w) \rangle_1 \\
&= (-1)^{[p] + [q]} \langle v | E_{pq} w \rangle_1'
\end{align*}
we see that the form $\langle ~ | ~ \rangle_1'$ obtained via the automorphism $\gamma$ satisfies the condition of the form $\langle ~ | ~ \rangle_2$ in equation (\ref{FormType1and2}). 
Furthermore, the forms $\langle ~ | ~ \rangle_1$ and $\langle ~ | ~ \rangle_2$ agree on the maximal $\mathbb{Z}$-graded component \cite{ZhaGou1990} of $V(\Lambda)$. 
Hence,  
$$
\langle v \vert w \rangle_1' = \langle v \vert w \rangle_2
$$
or equivalently
$$
\langle v \vert w \rangle_2 = \langle v \vert \gamma(w) \rangle_1.
$$
It follows that
\begin{align*}
\langle e^\Lambda_\beta \vert e^\Lambda_\alpha \rangle_1 &= (-1)^{(\beta)} \langle e^\Lambda_\beta \vert e^\Lambda_\alpha \rangle_2 \\
&= (-1)^{(\alpha)} \langle e^\Lambda_\beta \vert e^\Lambda_\alpha \rangle_2
\end{align*}
so that the two forms are equivalent up to a sign. Also, note that we follow the convention
$$
\lvert \langle v_0 | v_0 \rangle \rvert = 1
$$
where $v_0$ is the highest weight vector.

When considering coupling coefficients, we observe that the automorphism $\gamma$ acts like a group element on a tensor product space. We then have
\begin{align*}
\left\langle e^{\Lambda+\varepsilon_r}_\nu | e_\alpha \otimes e^\Lambda_\beta \right\rangle_2
&= \left\langle e^{\Lambda+\varepsilon_r}_\nu | \gamma(e_\alpha) \otimes \gamma(e^\Lambda_\beta) \right\rangle_1 \\
&= \left\langle \gamma(e^{\Lambda+\varepsilon_r}_\nu) | e_\alpha \otimes e^\Lambda_\beta \right\rangle_1 \\
&= (-1)^{(\nu)} \left\langle e^{\Lambda+\varepsilon_r}_\nu | e_\alpha \otimes e^\Lambda_\beta \right\rangle_1
\end{align*}
so that the coupling coefficients under the forms $\langle ~ | ~ \rangle_1$ and $\langle ~ | ~ \rangle_2$ are equivalent up to a sign. In subsequent sections we will assume $\langle ~ | ~ \rangle \equiv \langle ~ | ~ \rangle_1$ without loss of generality.

Note that the results of section \ref{SubSect_ContraDualWig} are given with respect to the form 
$\langle ~ | ~ \rangle_2$. The matrix elements
$$
\langle v^\Lambda_\beta \vert E_{pq} \vert v^\Lambda_\alpha \rangle_2 
$$
need to be multiplied by an overall phase $(-1)^{(\beta)}$ to give their values with respect to the form 
$\langle ~ | ~ \rangle_1$.

\subsection{General results}
We give our final results that are applicable for all dominant $\Lambda$ (including non-unitary highest weights) in terms of the unbarred roots $\alpha_k$ and ${\alpha_0}_k$. These expressions may be given in terms of the barred roots by a simple conversion. For example, the definitions $\alpha_i = \tilde{\Lambda}_i + m - n - i$ and $\bar{\alpha}_i = i - 1 - \tilde{\Lambda}_i$ imply that
$$
\bar{\alpha}_i = -\alpha_i + m - n - 1
$$
with a similar expression for the $\alpha_{0_i}$.

In terms of the following quantities
$$
\mu_r = \left[ \prod_{k\in \tilde{I},k\neq r} \left(\alpha_r - \alpha_k \right)^{-1}\prod_{k\in
I} \left(\alpha_r - {\alpha_0}_k - (-1)^{(k)}\right) \right]^{1/2}
,\ \ r\in \tilde{I} ,
$$
\begin{align*}
\mu_{r,u} &= (-1)^{(r)(u) +(r) + (u)} S(r-u)
\nn \\
&~~\times
\left[
\frac{
(-1)^{|I|}
\prod_{k\in I} \left(\alpha_r - {\alpha_0}_k - (-1)^{(k)}\right)
\prod_{k\in\tilde{I},k \neq r} \left(\alpha_k - {\alpha_0}_u \right)
}
{
({\alpha}_r-{\alpha}_u + 1)
\prod_{k\in \tilde{I},k\neq r} \left(\alpha_r - \alpha_k \right)
\prod_{k\in I,k\neq u} \left({\alpha_0}_u - {\alpha_0}_k - (-1)^{(k)}\right)
} 
\right]^{1/2}
,\ \ r\in \tilde{I},u\in \bar{I}_0 ,
\end{align*}
\begin{align*}
\mu_{r,u} &= (-1)^{(r)(u) +(r) + (u)} S(r-u) 
\nn\\
&~~\times
\left[
 (-1)^{|I|} \prod_{k\in \tilde{I},k\neq r} 
\left(
\frac 
{\alpha_k - {\alpha_0}_u  }
{\alpha_r - \alpha_k }
\right)
\prod_{k\in I,k \neq u} 
\left(
\frac
{ \alpha_r - {\alpha_0}_k - (-1)^{(k)} }
{{\alpha_0}_u - {\alpha_0}_k - (-1)^{(k)} }
\right)
\right]^{1/2}
,\ \ r\in \tilde{I},u\in I_1 ,
\end{align*}
the final expressions are
\begin{align*}
&\left\langle 
\begin{array}{c} \Lambda-\varepsilon_r\\ \lambda
 \end{array}
\right\rVert
\left.
\begin{array}{c} \bar{\varepsilon}_1 \\ 0 
 \end{array}
; \begin{array}{c} \Lambda \\ \lambda
 \end{array}
\right\rangle
= \mu_r, ~~r \in \tilde{I} \\
&\left\langle
\begin{array}{c} \Lambda-\varepsilon_r\\ \lambda-\varepsilon_{0_u} 
 \end{array}
\right\rVert
\left.
\begin{array}{c} \bar{\varepsilon}_1 \\ \bar{\varepsilon}_{0_1} 
 \end{array}
; \begin{array}{c} \Lambda \\ \lambda
 \end{array}
\right\rangle
= \mu_{r,u} ,~~r \in \tilde{I},u\in \bar{I}_0 \\
&\left\langle 
\begin{array}{c} \Lambda-\varepsilon_r\\ \lambda-\varepsilon_{0_u} 
 \end{array}
\right\rVert
\left.
\begin{array}{c} \bar{\varepsilon}_1 \\ \bar{\varepsilon}_{0_1} 
 \end{array}
; \begin{array}{c} \Lambda \\ \lambda
 \end{array}
\right\rangle 
= \mu_{r,u} ,~~r \in \tilde{I},u\in I_1 \\
&\left\langle 
\begin{array}{c} \Lambda+\varepsilon_r\\ \lambda
 \end{array}
\right\rVert
\left.
\begin{array}{c} \varepsilon_1 \\ 0 
 \end{array}
; \begin{array}{c} \Lambda \\ \lambda
 \end{array}
\right\rangle =  \mu_r ,~~ r\in \tilde{I}' \\
 &\left\langle
\begin{array}{c} \Lambda+\varepsilon_r\\ \lambda+\varepsilon_{0_u} 
 \end{array}
\right\rVert
\left.
\begin{array}{c} \varepsilon_1 \\ \varepsilon_{0_1} 
 \end{array}
; \begin{array}{c} \Lambda \\ \lambda
 \end{array}
\right\rangle = (-1)^{(r) + (u)} \mu_{r,u},~~ r\in \tilde{I'},u\in I_0 \\
 &\left\langle
\begin{array}{c} \Lambda+\varepsilon_r\\ \lambda+\varepsilon_{0_u} 
 \end{array}
\right\rVert
\left.
\begin{array}{c} \varepsilon_1 \\ \varepsilon_{0_1} 
 \end{array}
; \begin{array}{c} \Lambda \\ \lambda
 \end{array}
\right\rangle = (-1)^{(r) + (u)} \mu_{r,u},~~ r\in \tilde{I'},u\in I_1 .
\end{align*}
These formulae give RWCs in the case that $V\otimes V(\Lambda)$ and $V^*\otimes V(\Lambda)$ are
completely reducible. Note that the formulae obtained, however, may still
be applicable more generally. 
Finally, the full WCs are given in terms of the RWCs in equations (\ref{FullWC1}) and (\ref{FullWC2}).
\subsection{Symmetry relations}
In this section we assume that
$$
V \otimes V(\Lambda),~V^* \otimes V \otimes V(\Lambda) 
$$
are completely reducible.
Consider a $gl(m|n)$ vector operator $\psi^i$. Acting on an irreducible module $V(\Lambda)$, $\psi^i$ determines an intertwining operator 
$$
\psi: V \otimes V(\Lambda) \rightarrow W, ~~ \psi(e_i \otimes v) = \psi^i v, ~~\forall v \in V(\Lambda)
$$
where $W$ is the direct sum of the irreducible modules
$$
W = \bigoplus_r V(\Lambda + \varepsilon_r),
$$
where each component of the direct sum on the right hand side is interpreted to vanish identically
if the corresponding $\Lambda+\varepsilon_r$ is not dominant.
From Schur's lemma we have
\begin{align*}
\left\langle e_\beta^{\Lambda  + \varepsilon_r} | \psi^i | e^{\Lambda}_\alpha \right\rangle 
= \left \langle \Lambda + \varepsilon_r || \psi || \Lambda \right \rangle
\left\langle e^{\Lambda+\varepsilon_r}_\beta | e_i \otimes e^{\Lambda}_\alpha \right\rangle
\end{align*}
which is the Wigner-Eckart theorem for vector operators. Similarly, for a dual vector operator $\phi_i$ defined by
$$
\phi_i = (\psi^i)^\dagger
$$
we have
\begin{align*}
\left\langle e_\alpha^\Lambda \vert \phi_i \vert e^{\Lambda  + \varepsilon_r}_\beta \right\rangle 
= \left \langle \Lambda || \phi || \Lambda + \varepsilon_r \right \rangle
\left\langle e^\Lambda_\alpha | e_i \otimes e^{\Lambda + \varepsilon_r}_\beta \right\rangle
\end{align*}
which implies
\begin{align*}
&\left \langle \Lambda + \varepsilon_r || \psi || \Lambda \right \rangle
\left\langle e^{\Lambda+\varepsilon_r}_\beta | e_i \otimes e^{\Lambda}_\alpha \right\rangle \\
&~~= \left\langle e_\beta^{\Lambda  + \varepsilon_r} | \psi^i | e^{\Lambda}_\alpha \right\rangle \\
&~~= \left\langle e_\alpha^{\Lambda} | \phi_i | e^{\Lambda + \varepsilon_r}_\beta \right\rangle^*\\
&~~=\left \langle \Lambda || \phi || \Lambda + \varepsilon_r \right \rangle^*
\left\langle e^{\Lambda}_\alpha | \bar{e}_i \otimes e^{\Lambda+\varepsilon_r}_\beta \right\rangle^*.
\end{align*}
Thus we arrive at the symmetry relation
\begin{align}
\label{SymRel}
\left\langle e^{\Lambda+\varepsilon_r}_\beta | e_i \otimes e^{\Lambda}_\alpha \right\rangle
&= \eta \left\langle e^{\Lambda}_\alpha | \bar{e}_i \otimes e^{\Lambda+\varepsilon_r}_\beta \right\rangle^* \nn\\
&= \eta \left\langle \bar{e}_i \otimes e^{\Lambda+\varepsilon_r}_\beta |  e^{\Lambda}_\alpha \right\rangle
\end{align}
where $\eta$ is a constant that is independent of $i, \alpha$ and $\beta$ and is given by
$$
\left \langle \Lambda + \varepsilon_r || \psi || \Lambda \right \rangle \eta = \left \langle \Lambda
|| \phi || \Lambda + \varepsilon_r \right \rangle^*.
$$
Even when $V^* \otimes V \otimes V(\Lambda)$ is not completely reducible we may still use the above symmetry condition as a definition of the covariant dual vector Wigner coefficients.
We will now determine $\eta$ (up to a phase) from the expression
\begin{align}
\eta^2 = \frac {\bar{\delta}_r [\Lambda + \varepsilon_r]} {\delta_r [\Lambda] } 
\label{eta_ratio}
\end{align}
where $\bar{\delta}_r [\Lambda + \varepsilon_r]$ indicates that a shift is applied to the $\bar{\delta}_r$ expression. 
To aid the calculation of $\eta$ we use the index set free version of $\bar{\delta}_u$ and $\delta_u$  given in \cite{GIW2,GIW3}
\begin{align}
\delta_u [\Lambda] &= \prod_{k = 1, \neq u}^m \left(\frac{\alpha_k-\alpha_u}{\alpha_u - \beta_k - 1}  \right)
\prod_{k=m+1,\neq u}^{m+n}(\alpha_u-\alpha_k + 1 )^{-1}
\prod_{k=m+1}^{m+n+1}(\beta_k-\alpha_u ),\ \ u \in I',
\label{delta_dontshift}
\end{align}
\begin{align}
\bar{\delta}_u [\Lambda] &= \prod_{k = 1}^m \left(\frac{\alpha_k-\alpha_u + 1 - (-1)^{(u)}}{\beta_k-\alpha_u + 1 - (-1)^{(u)}}  \right)
\prod_{k=m+1}^{m+n+1}(\beta_k-\alpha_u-(-1)^{(u)})
\prod_{k=m+1,\neq u}^{m+n}(\alpha_u-\alpha_k+(-1)^{(u)})^{-1},\ \ u \in I.
\label{deltabar_unshifted}
\end{align}
From the definitions of the roots $\alpha_i$ and $\alpha_\mu$ we see that 
the shift in the weight label $\Lambda_u \rightarrow \Lambda_u - 1$ is equivalent to a shift in 
the roots $\alpha_u \rightarrow \alpha_u - (-1)^{(u)}$. Applying this shift to equation (\ref{deltabar_unshifted}) we have
\begin{align}
\bar{\delta}_u [\Lambda - \varepsilon_u] &= \prod_{k = 1}^m \left(\frac{\alpha_k-\alpha_u + 1 }{\beta_k-\alpha_u + 1 }  \right)
\prod_{k=m+1}^{m+n+1}(\beta_k-\alpha_u)
\prod_{k=m+1,\neq u}^{m+n}(\alpha_u-\alpha_k)^{-1},\ \ u \in I.
\label{deltabar_shifted}
\end{align}
Substituting equations (\ref{deltabar_shifted}) and (\ref{delta_dontshift}) into equation (\ref{eta_ratio}) gives
\begin{align}
\eta^2 &= \prod_{k=1}^m \frac {\alpha_k - \alpha_u + 1}{\alpha_k - \alpha_u}
  \prod_{k=m+1}^{m+n} \frac {\alpha_k - \alpha_u - 1}{\alpha_k - \alpha_u} \nn\\
&= \prod^{m+n}_{k=1} 
\left( 
\frac
{\alpha_u - \alpha_k - (-1)^{(k)}}
{\alpha_u - \alpha_k}
\right) \\
&= \frac {D(\Lambda - \varepsilon_u)} { D(\Lambda) } \nn,
\end{align}
where $D(\Lambda)$ is the dimension of the even $gl(m) \oplus gl(n)$ submodule of the $gl(m|n)$ module of highest weight $\Lambda$. Noting that the RMEs have phase $+1$ we obtain the remarkably nice result
\begin{align}
\label{SymRelConst}
\eta = \sqrt { \frac {D(\Lambda - \varepsilon_u)} { D(\Lambda) } },
\end{align}
where the positive square root is taken.

\subsection{Concluding remarks}
In this article we have given closed form expressions for the covariant vector
and contravariant dual vector Wigner coefficients. These results were shown to
be valid independently of the form chosen. The symmetry condition also allows
a direct determination of the contravariant vector and covariant dual vector Wigner coefficients when 
 $V^* \otimes V \otimes V(\Lambda)$ is completely reducible.

It is an interesting question to determine the highest weights $\Lambda$ for which 
 $V^* \otimes V \otimes V(\Lambda)$ is not reducible. These would include the atypical mixed tensor representations.
 
Finally, the symmetry relation (\ref{SymRel}) is particularly remarkable in that the proportionality
constant in equation (\ref{SymRelConst}) avoids the occurrence of zero superdimensions that occur in
the superalgebra case.


%
%
%
\section*{Acknowledgements}
This work was supported by the Australian Research Council through Discovery Project
DP140101492. 
%
%
%
%
%

\end{document}